\documentclass[preprint2]{aastex61}

\received{June 17, 2020}
\revised{August 21, 2020}
\accepted{August 26, 2020}
\shorttitle{The two-component ejecta of SN 2016gkg}
\shortauthors{Kuncarayakti et al.}
\usepackage[caption=false]{subfig}

\begin{document}

\title{Direct evidence of two-component ejecta in supernova 2016gkg from nebular spectroscopy\footnote{Based on observations collected at the European Organisation for Astronomical Research in the Southern Hemisphere under ESO program 0100.D-0341; 
the Gemini Observatory, which is operated by the Association of Universities for Research in Astronomy, Inc., under a cooperative agreement with the NSF on behalf of the Gemini partnership: the National Science Foundation (United States), National Research Council (Canada), CONICYT (Chile), Ministerio de Ciencia, Tecnolog\'{i}a e Innovaci\'{o}n Productiva (Argentina), Minist\'{e}rio da Ci\^{e}ncia, Tecnologia e Inova\c{c}\~{a}o (Brazil), and Korea Astronomy and Space Science Institute (Republic of Korea), programs GS-2017B-Q-65, GS-2018B-FT-107 (Fast Turnaround); 
and the Subaru Telescope, which is operated by the National Astronomical Observatory of Japan, program S17B-162S.}}

\correspondingauthor{Hanindyo Kuncarayakti}
\email{hanindyo.kuncarayakti@utu.fi}

\author[0000-0002-1132-1366]{Hanindyo Kuncarayakti}
\affil{Tuorla Observatory, Department of Physics and Astronomy, FI-20014 University of Turku, Finland}
\affil{Finnish Centre for Astronomy with ESO (FINCA), FI-20014 University of Turku, Finland}

\author{Gast\'on Folatelli}
\affiliation{Facultad de Ciencias Astron\'omicas y Geof\'isicas, Universidad Nacional de La Plata, Paseo del Bosque S/N, B1900FWA La Plata, Argentina}
\affiliation{Instituto de Astrof\'isica de La Plata (IALP), CONICET, Argentina}
\affiliation{Kavli Institute for the Physics and Mathematics of the Universe (WPI), The University of Tokyo, 5-1-5 Kashiwanoha, Kashiwa, Chiba 277-8583, Japan}
%\collaboration{(AAS Journals Data Scientists collaboration)}

\author{Keiichi Maeda}
\affiliation{Department of Astronomy, Graduate School of Science, Kyoto University, Sakyo-ku, Kyoto 606-8502, Japan}
\affiliation{Kavli Institute for the Physics and Mathematics of the Universe (WPI), The University of Tokyo, 5-1-5 Kashiwanoha, Kashiwa, Chiba 277-8583, Japan}
%\nocollaboration

\author{Luc Dessart}
\affiliation{Institut d'Astrophysique de Paris, CNRS-Sorbonne Universit\'e, 98 bis boulevard Arago, F-75014 Paris, France}

\author{Anders Jerkstrand}
\affiliation{Max-Planck-Institut f\"ur Astrophysik, Karl-Schwarzschild Str 1, D-85748 Garching, Germany}
\affiliation{Department of Astronomy, Stockholm University, The Oskar Klein Centre, AlbaNova, SE-106 91 Stockholm, Sweden}

\author{Joseph P. Anderson}
\affiliation{European Southern Observatory, Alonso de C\'ordova 3107, Casilla 19, Santiago, Chile}

\author{Kentaro Aoki}
\affiliation{Subaru Telescope, National Astronomical Observatory of Japan, 650 North A'ohoku Place, Hilo, HI 96720, USA}

\author{Melina C. Bersten}
\affiliation{Facultad de Ciencias Astron\'omicas y Geof\'isicas, Universidad Nacional de La Plata, Paseo del Bosque S/N, B1900FWA La Plata, Argentina}
\affiliation{Instituto de Astrof\'isica de La Plata (IALP), CONICET, Argentina}
\affiliation{Kavli Institute for the Physics and Mathematics of the Universe (WPI), The University of Tokyo, 5-1-5 Kashiwanoha, Kashiwa, Chiba 277-8583, Japan}

\author{Luc\'ia Ferrari}
\affiliation{Facultad de Ciencias Astron\'omicas y Geof\'isicas, Universidad Nacional de La Plata, Paseo del Bosque S/N, B1900FWA La Plata, Argentina}

\author{Llu\'is Galbany}
\affiliation{Departamento de F\'isica Te\'orica y del Cosmos, Universidad de Granada, E-18071 Granada, Spain}

\author{Federico Garc\'ia}
\affiliation{Kapteyn Astronomical Institute, University of Groningen, PO BOX 800, Groningen NL-9700 AV, the Netherlands}

\author{Claudia P. Guti\'errez}
\affiliation{Department of Physics and Astronomy, University of Southampton, Southampton, SO17 1BJ, UK}

\author{Takashi Hattori}
\affiliation{Subaru Telescope, National Astronomical Observatory of Japan, 650 North A'ohoku Place, Hilo, HI 96720, USA}

\author{Koji S. Kawabata}
\affiliation{Hiroshima Astrophysical Science Center, Hiroshima University, 1-3-1 Kagamiyama, Higashi-Hiroshima, Hiroshima 739-8526, Japan}

\author{Timo Kravtsov}
\affil{Tuorla Observatory, Department of Physics and Astronomy, FI-20014 University of Turku, Finland}

\author{Joseph D. Lyman}
\affiliation{Department of Physics, University of Warwick, Coventry CV4 7AL, UK}

\author{Seppo Mattila}
\affil{Tuorla Observatory, Department of Physics and Astronomy, FI-20014 University of Turku, Finland}

\author{Felipe Olivares E.}
\affiliation{Instituto de Astronom\'{\i}a y Ciencias Planetarias, Universidad de Atacama, Copayapu 485, Copiap\'o, Chile}
%\affiliation{Millennium Institute of Astrophysics (MAS), Nuncio Monse\~nor S\'otero Sanz 100, Providencia, Santiago, Chile}

\author{Sebasti\'an F. S\'anchez}
\affiliation{Instituto de Astronom\'ia, Universidad Nacional Aut\'onoma de M\'exico, Circuito Exterior, Ciudad Universitaria, Ciudad de M\'exico 04510, Mexico}

\author{Schuyler D. Van Dyk}
\affiliation{Caltech/IPAC, Pasadena, CA 91125, USA}

%\altaffiliation{Creator of AASTeX v6.1}
%\collaboration{(LaTeX collaboration)}
%\affiliation{AAS Director of Publishing}
%\affiliation{American Astronomical Society \\
%2000 Florida Ave., NW, Suite 300 \\
%Washington, DC 20009-1231, USA}
%
%\author{Jeff Lewandowski}
%\affiliation{IOP Senior Publisher for the AAS Journals}
%\affiliation{IOP Publishing, Washington, DC 20005}



\begin{abstract}

%This example manuscript is intended to serve as a tutorial and template for
%authors to use when writing their own AAS Journal articles. The manuscript
%includes a history of \aastex\ and documents the new features in the
%previous version, 6.0, as well as the new features in version 6.1. This
%manuscript includes many figure and table examples to illustrate these new
%features.  Information on features not explicitly mentioned in the article
%can be viewed in the manuscript comments or more extensive online
%documentation. Authors are welcome replace the text, tables, figures, and
%bibliography with their own and submit the resulting manuscript to the AAS
%Journals peer review system.  The first lesson in the tutorial is to remind
%authors that the AAS Journals, the Astrophysical Journal (ApJ), the
%Astrophysical Journal Letters (ApJL), and Astronomical Journal (AJ), all
%have a 250 word limit for the abstract.  If you exceed this length the
%Editorial office will ask you to shorten it.

  {Spectral observations of the type-IIb supernova (SN) 2016gkg {at 300-800 days} are reported. The {spectra show} nebular characteristics, revealing emission from the progenitor star's metal-rich core and providing clues to the kinematics and physical conditions of the explosion. The nebular spectra are dominated by emission lines of [O~I]~$\lambda\lambda6300, 6364$ and [Ca~II]~$\lambda\lambda7292, 7324$. Other notable, albeit weaker, emission lines include Mg~I]~$\lambda4571$, [Fe~II]~$\lambda7155$, O~I~$\lambda7774$, Ca II triplet, and a broad, boxy feature at the location of H$\alpha$. Unlike in other stripped-envelope SNe, the [O~I] doublet is clearly resolved due to the presence of strong narrow components. The doublet shows an unprecedented emission line profile consisting of at least three components for each [O~I]$\lambda6300, 6364$ line: a broad component (width $\sim2000$ km~s$^{-1}$), and a pair of narrow blue and red components (width $\sim300$ km~s$^{-1}$) mirrored against the rest velocity. The narrow component appears also in other lines, and is conspicuous in [O~I]. This indicates the presence of multiple distinct kinematic components of material at low and high velocities. The low-velocity components are likely to be produced by a dense, slow-moving emitting region near the center, while the broad components are emitted over a larger volume. These observations suggest an asymmetric explosion, supporting the idea of two-component ejecta that influence the resulting late-time spectra and light curves. SN~2016gkg thus presents striking evidence for significant asymmetry in a standard-energy SN explosion. The presence of material at low velocity, which is not predicted in 1D simulations, emphasizes the importance of multi-dimensional explosion modeling of SNe.
   }

\end{abstract}

\keywords{supernovae: general --- 
supernovae: individual (SN~2016gkg) --- stars: massive --- stars: interiors}



\section{Introduction}

{Type-IIb supernovae (SNe IIb) exhibit} a change of spectral appearance from showing hydrogen lines at early times, to a typical type-Ib SN (helium-rich, hydrogen-poor), as first exemplified by SN 1987K \citep{filippenko88} and SN 1993J \citep{woosley94}. The progenitor stars of this subclass are thought to retain a thin layer of hydrogen at the time of the explosion, giving rise to the SN II $\rightarrow$ SN Ib spectral transition.

SN~2016gkg is arguably one of the best-observed members of this subclass, with the detection of the progenitor star in pre-explosion archival images \citep{kilpatrick17,tartaglia17} and the extraordinary discovery of the optical light breakout within two hours of core collapse \citep{bersten18}. The early light curve of SN~2016gkg following the shock breakout is exceptionally well sampled, covering the peak associated with the shock breakout cooling, {as well as} the subsequent main peak resulting from radioactive heating {that is} frequently observed in other SNe. 
The post shock cooling peak of the light curve is critical to constrain the progenitor radius and the mass of its extended envelope {\citep[see e.g.][for a general review, and references therein]{waxman17}.} 
With this method, the radius {of the progenitor of SN 2016gkg} has been constrained to be between $\sim$200-300 $R_\odot$ and the extended envelope mass in the order of $10^{-2}$ $M_\odot$ \citep{piro17,bersten18}.
{\citet{arcavi17} reported a similar envelope mass estimate, albeit with a smaller radius of 40-150 $R_\odot$.}
The detection of the  pre-explosion progenitor star candidate in {Hubble Space Telescope (HST)} archival images points to an initial mass of 15-20 $M_\odot$ \citep{kilpatrick17,tartaglia17}.
\citet{bersten18} further improved the identification and characterization of the progenitor star in the {HST} images, and concluded that the progenitor may have been a 19.5 $M_\odot$ star in a binary system via binary star evolution modeling.

Here we report late-time spectral observations of SN~2016gkg, resulting in spectra between +300-800 days after the light curve maximum.
At this late stage, the SN ejecta have expanded and become optically thin, rendering the emission from the deeper layers and inner core visible. 
As will be demonstrated in the following sections, this condition enables the detection of very slow-moving material, in addition to the typically seen fast-moving material, in the ejecta of SN~2016gkg, which uniquely characterizes the object.
The observations and data reduction are described in the next section, followed by discussions of the results and interpretations.
{Herewith ``light curve maximum'' refers to the peak in the light curve that resulted from radioactive heating, which occurred on MJD 57668.4 or 2016~October~7.4 (UT), in {B} and {V} bands \citep{bersten18}.}

\section{Observations and data reductions}
\label{sec:obs}

SN~2016gkg was observed with the Gemini South telescope and the GMOS-S instrument \citep{hook04,gimeno16}, on 2017 August 20 and 26 (UT dates are used throughout). At the median date of observation, August 23, the phase corresponds to +320~d after light curve maximum. The observations were conducted using the longslit mode with the R400 grism, with a $4\times4$ binning and a total exposure time of 5.3~h. 
Another GMOS-S longslit observation was performed on 2018 November 10 (+764~d) through the Gemini Fast Turnaround program, using the R400 grism and $4\times2$ binning, with a total exposure time of 3.0~h.
Spectral dithering was performed to eliminate cosmic rays and cover the gaps between the GMOS CCDs.
Data reduction was done using the Gemini package in IRAF\footnote{IRAF is distributed by the National Optical Astronomy Observatory, which is operated by the Association of Universities for Research in Astronomy, Inc., under cooperative agreement with the National Science Foundation.}.
All the spectra were wavelength-calibrated, and flux-calibrated using spectrophotometric standard stars.

On 2017 November 30, the field of SN~2016gkg was observed using the MUSE instrument \citep{bacon14} at the ESO Very Large Telescope. MUSE was used in Wide Field Mode, with a total exposure time of 2400~s. SN~2016gkg was positioned at the center of the field of view and the observation captured the 1'$\times$1' field surrounding SN~2016gkg in integral field spectroscopy, which includes the nebular spectrum of the object at +419~d past maximum. 
The MUSE spaxel size {is}  0.2"$\times$0.2", corresponding to $\sim25 \times 25$ pc at the assumed distance of 26.4 Mpc to SN~2016gkg \citep{bersten18}.
The observations were part of the All-weather MUSe Supernova Integral-field Nearby Galaxies \citep[AMUSING; see][]{galbany16} survey.
MUSE data reduction was performed using MUSE data reduction pipelines run under ESO Reflex \citep{freudling13}. The datacube was sky-subtracted using blank sky pointings, and corrected for atmospheric effects using Zurich Atmospheric Package \citep[ZAP;][]{soto16}. The measured image quality in the final datacube is typically 0.8" FWHM (full width at half maximum) across the wavelengths, and this value is used as the radius of the circular aperture for 1-D spectrum extraction of SN~2016gkg with QFitsView \citep{ott12}.

SN~2016gkg was also observed using FOCAS \citep{kashikawa02} at the Subaru telescope on 2018 January 1 (+451~d) in the 0\arcsec.8 off-center slit mode with the B300 grism + Y47 order-cut filter setting. 
The total exposure time was 2400~s, taken in two exposures of 1200 s each, with a slight positional shift (AB pattern). 
%The spectrum was wavelength calibrated with a ThAr lamp, with small shifts by cross-correlating with the sky emission lines, and flux calibration with Feige 110. 
Standard data reduction procedure which includes wavelength and flux calibration was done using IRAF, with cosmic ray removal using L.A.Cosmic \citep{vandokkum01}.

The final reduced spectra from these observations were then corrected for the redshift of $z = 0.0049$ as measured from the underlying H~II region emission lines, that agrees with the reported redshift of the host galaxy NGC 613 \citep{meyer04}.
This redshift is assumed to be the rest reference of the SN throughout the paper.

The spectral resolutions measured from the sky emission lines in the spectra are $\sim$3~\AA~FWHM for MUSE (corresponding to resolving power $R\approx2200$ and velocity resolution $\Delta v \approx 140$ km~s$^{-1}$ at H$\alpha$ wavelength), and $\sim$10~\AA~for GMOS and FOCAS ($R \approx 700$ and $\Delta v \approx 430$ km s$^{-1}$).

\section{Results and discussions}

\subsection{Spectrum and line identifications}

\begin{figure*}
   \centering
   \includegraphics[width=\hsize]{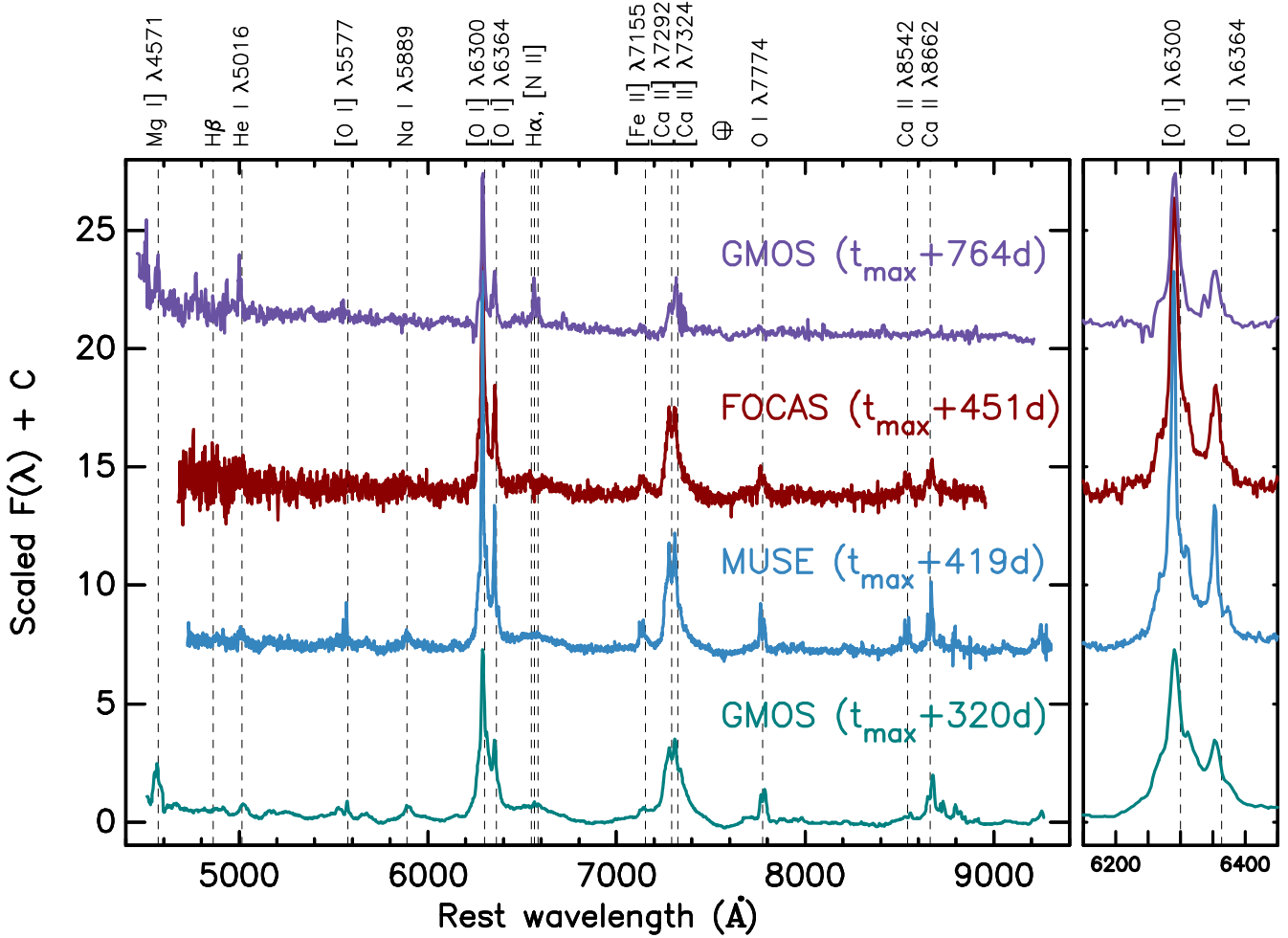}
\caption{Nebular spectra of SN~2016gkg. The species responsible for the emission lines are shown, with their rest wavelengths indicated by vertical dashed lines. 
Narrow H$\alpha$+[N~II] are likely lines arising in the interstellar medium (ISM); telluric absorption region is indicated with the Earth symbol ($\Earth$, $\sim$7600 \AA).
Fluxes have been scaled to allow comparison between the spectra. A zoom-in on the [O I] doublet is shown on the right-hand panel inset.
Line identification throughout the paper follows that of \citet{groningsson08} and \citet{jerkstrand15}.
{The spectra are available as the Data behind the Figure (DbF).}
}
\label{specall}
\end{figure*}

\begin{figure}
   \centering
   \includegraphics[width=\hsize]{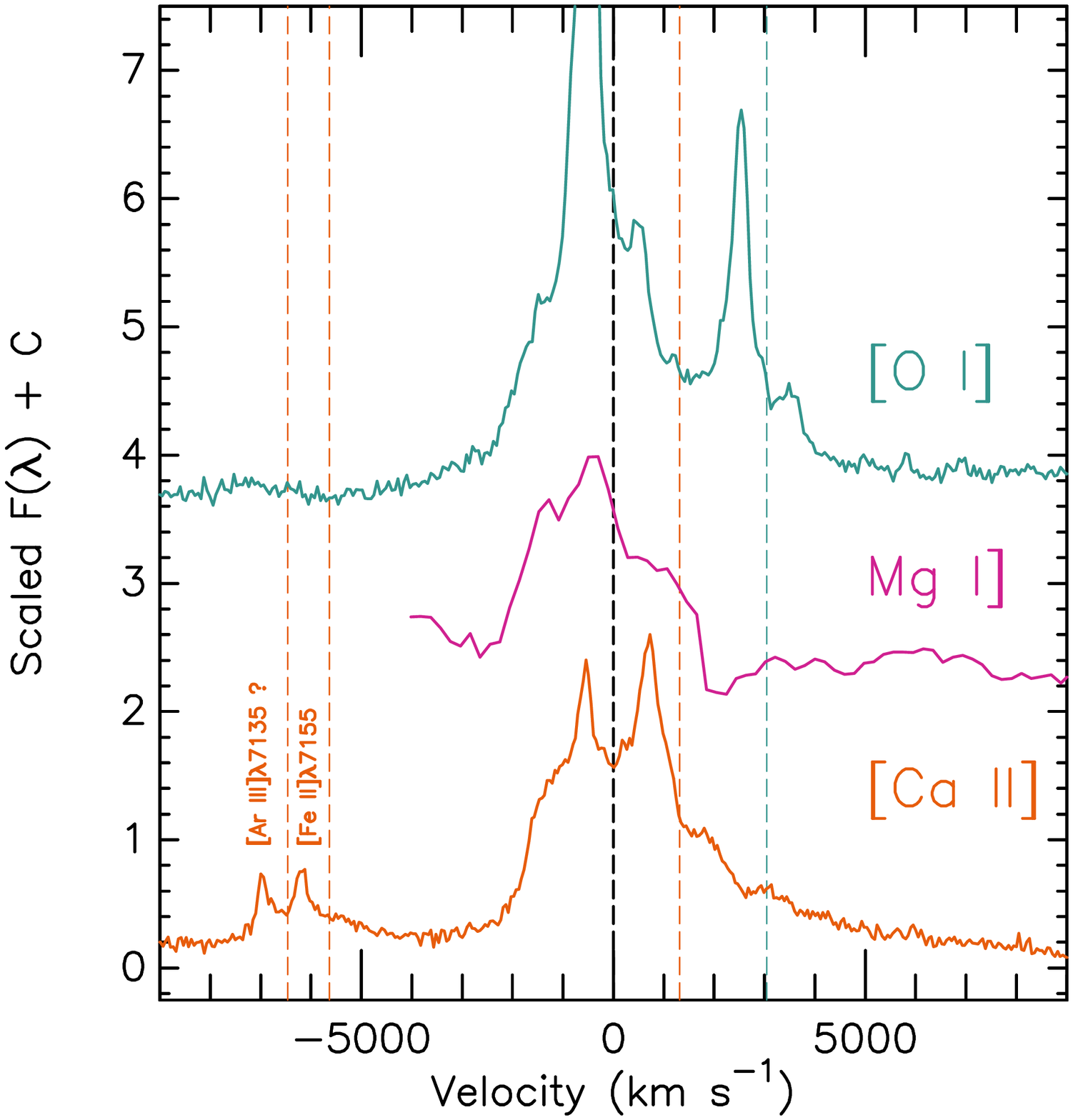}
\caption{[O~I] $\lambda$6300 (teal), Mg~I] $\lambda$4571 (magenta), and [Ca~II] $\lambda$7292 (orange) lines in velocity space. The Mg~I] line is from the +320~d spectrum, while the others are from the +419~d one.
Black vertical dashed line indicates the zero velocity for each individual line, and colored dashed lines correspond to the rest wavelengths of [O~I] $\lambda$6364, [Ca~II] $\lambda$7324, additionally [Fe~II] $\lambda7155$ and possibly [Ar~III] $\lambda7135$, seen near [Ca~II].}
\label{specvel}
\end{figure}

The four nebular spectra of SN~2016gkg at +320, +419, +451 and +764~d are presented in Figure~\ref{specall}. 
{These epochs are considerably late in the SN evolution as most stripped-envelope (SE) SNe enter the nebular phase as early as few months after maximum. The majority of nebular spectra available in the literature correspond to phases earlier than one year post maximum light, while only few SESNe have been observed at the phase of one year or later.}
Typical for a hydrogen-poor core collapse SN observed at {the} nebular late-phase stage, the spectra {of SN~2016gkg} show broad nebular emission lines, dominated by the [O~I] $\lambda\lambda$6300,6364 doublet and [Ca~II] $\lambda\lambda$7292,7324 \citep[e.g.][]{maeda08,tauben09,fang19}. 
The [O~I] $\lambda\lambda$6300,6364 doublet shows an extraordinary line profile with strong narrow emission components. These narrow components (FWHM $\sim$ 300 km~s$^{-1}$) are fully resolved with the highest resolution spectrum from MUSE ($\Delta v \approx 140$ km~s$^{-1}$).
The [Ca~II] $\lambda\lambda$7292,7324 line exhibit a sloping red wing {(see Figure~\ref{specall}, and a zoom-in on Figure~\ref{specvel})}, possibly arising from contamination by [Ni~II] $\lambda$7377 \citep{jerkstrand15}, or opacity and scattering effects \citep{jerkstrand17}.
Other weaker emission lines such as Mg~I], He~I, Na~I, Fe~II are also present, as well as a few lines from permitted transitions of O~I and Ca~II. 
The spectral lines are largely symmetric and no asymmetries that might have resulted from newly formed dust in the ejecta are detected \citep[see e.g.][]{meikle11}.

\begin{figure}[h]
   \centering
   \includegraphics[width=\hsize]{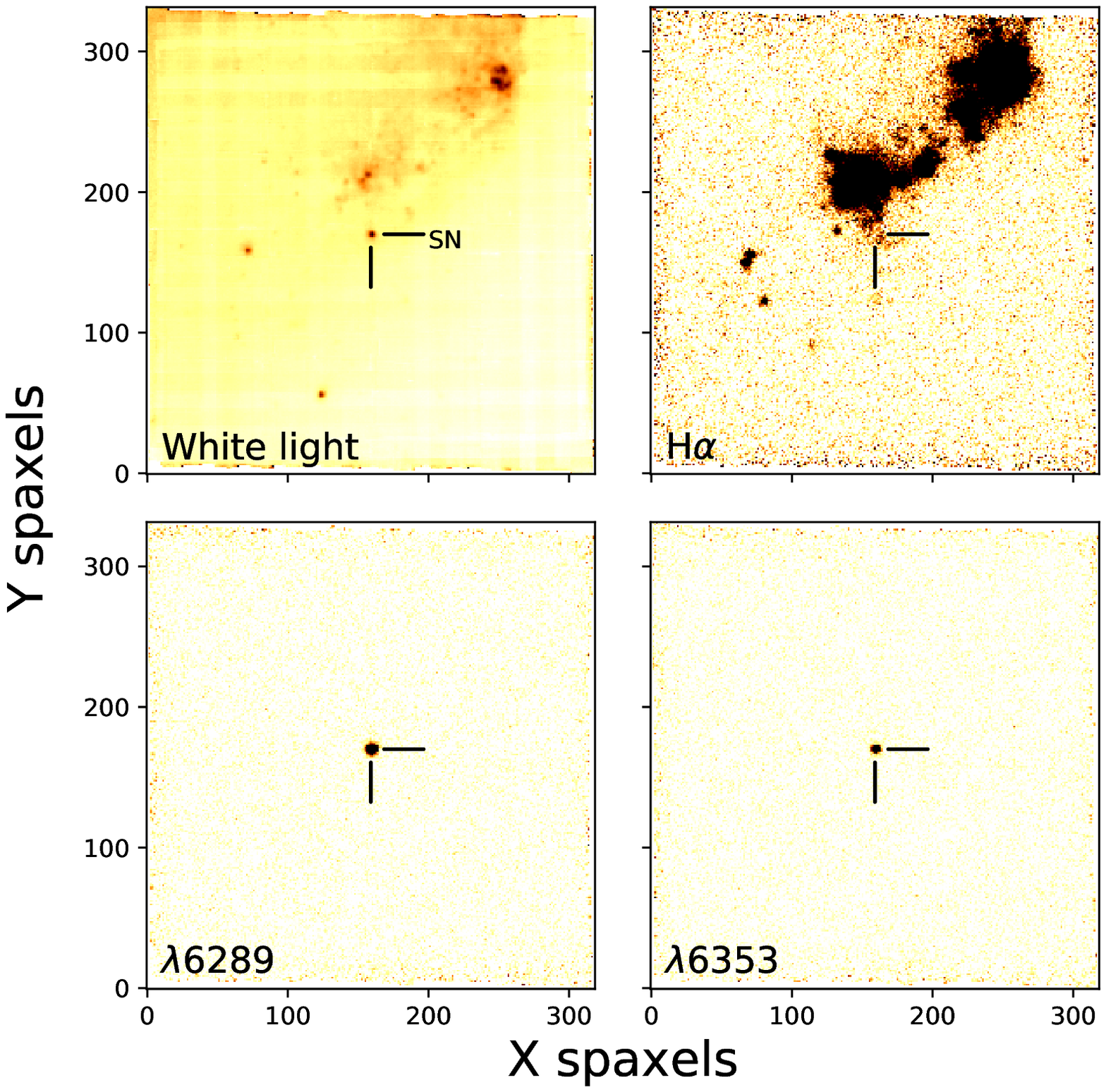}
\caption{Images of the SN environment in the host galaxy, generated from the MUSE datacube, spanning around 1~arcmin (300 spaxels) on each side. The SN position is indicated with crosshairs, and images are oriented north-up and east-left. 
The host galaxy center is off the field, due north.
Clockwise from top left: integrated (white) light, and narrow emission line maps of H$\alpha$ (6563 \AA), [O~I]$\lambda6364$ (at 6353 \AA), [O~I]$\lambda6300$ (at 6289 \AA). Emission line maps are continuum subtracted.
}
\label{mapfig}
\end{figure}

The narrow emission lines are checked by constructing linemaps from the MUSE datacube at their respective wavelengths. All the detected lines in the SN spectrum are confirmed to be emitted by the SN, as opposed to having originated from the diffuse background or noise, from the appearance of the SN as a point source in each linemap (Figure~\ref{mapfig}). This is one advantage of integral field spectroscopy.

\begin{figure}
   \centering
   \includegraphics[width=\hsize]{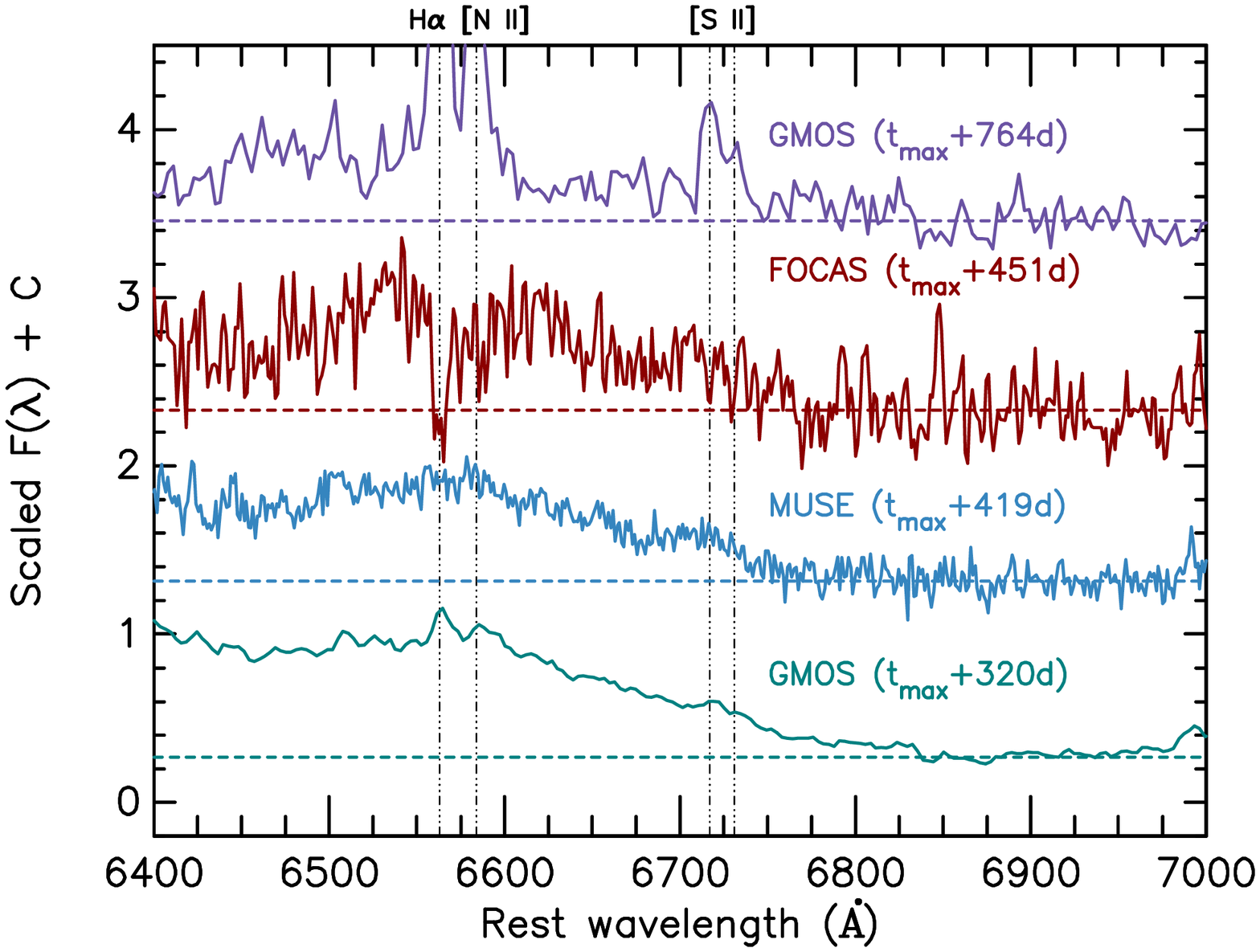}
\caption{{Spectra of SN~2016gkg around the H$\alpha$ region. Dashed horizontal lines indicate the average continuum level between 6850 and 6900 \AA. Likely host galaxy ISM lines of H$\alpha$, [N~II] $\lambda$6584, and [S~II] $\lambda$6717,6731 are indicated with vertical dash-dotted lines. These lines are clearly visible in emission in the +764 d spectrum, and appear as background subtraction artifacts in the +451 d spectrum.}}
\label{specpl_Ha}
\end{figure}

There is a relatively weak bump with a broad, boxy profile at the wavelength of H$\alpha$ in the spectra {(Figure~\ref{specpl_Ha})}, typical of SNe IIb at late times \citep[see e.g.][]{fang19}. 
{The bump is visible in all epochs.}
This bump has been suggested to be contributed by [N~II] $\lambda\lambda$6548,6583 \citep{jerkstrand15}, although CSM interaction may also give rise to {a boxy, flat-topped H$\alpha$ emission} as seen in the case of the type-IIb SN~1993J \citep{matheson00}.
A slight blueshift in the SN emission line peaks with respect to the rest wavelengths ($\sim -500$ km s$^{-1}$) can be discerned (also see Figure~\ref{specvel}). This phenomenon appears to be commonplace in SESNe, and may be explained by various factors such as ejecta geometry and opacity effects (\citealt{tauben09}; also see radiative transfer models presented in \citealt{dessart05}).

There is generally little evolution in the spectrum of SN~2016gkg between +320 to +450 d and the narrow+broad line profile persists {(Figure~\ref{specall})}.
Despite the different spectral resolutions, all spectra visibly display nearly identical sets of emission lines. 
The last spectrum at +764~d shows that most of the weaker lines have disappeared {beyond our detection limit}, leaving mostly the [O~I] and [Ca~II] doublets and Mg~I].
Narrow H$\alpha$ emission line accompanied by [N~II] $\lambda$6584 and the [S~II] $\lambda\lambda$6717,6731 doublet are present {(Figure~\ref{specpl_Ha})}, but these are likely to be lines originating from the background interstellar medium (ISM). These lines are better subtracted in the other spectra, and their linemaps do not reveal a point source at the SN position (Figure~\ref{mapfig}).

\subsection{Line profiles and comparison to other SNe}

The [O~I] $\lambda\lambda$6300,6364 doublet in the spectra of SN~2016gkg shows a narrow component superposed on a broad component, and while less prominent, such {composite} profile is also seen in [Ca~II] $\lambda\lambda$7292,7324 {(Figure \ref{specvel})}.
Comparing with available spectra in the literature, the line profile containing clear broad+narrow components of [O~I] and [Ca~II] in SN~2016gkg is unprecedented. Most SNe show line profiles of either a single Gaussian, double peak, or asymmetric/multipeaked, while those with a narrow core like seen in SN~2016gkg are relatively rare \citep{maeda08,tauben09} and the flux associated with the narrow core is never seen dominating over the broad component.

\begin{figure*}
   \centering
   \includegraphics[width=\hsize]{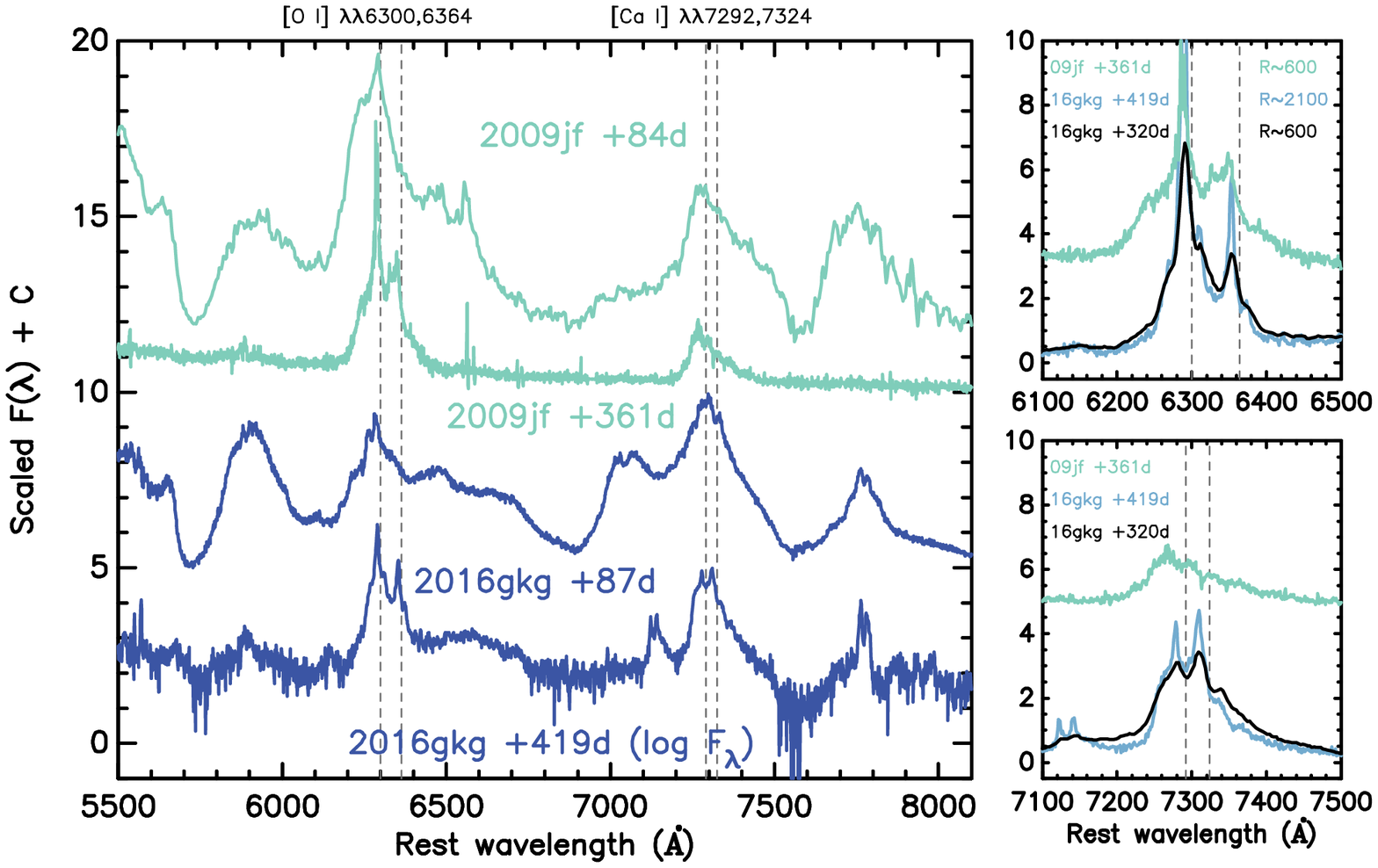}
\caption{{{(Left panel)}} Emergence of the [O~I] narrow line core in SN~2016gkg and SN~2009jf. Early-phase spectra of the two SNe are from \citet{bersten18} and \citet{valenti11}, respectively.
The +419~d spectrum of SN~2016gkg is plotted with the ordinate in log scale due to the extreme intensity of the narrow component.
{{(Right panels)} Comparison of the +361 d spectrum of SN~2009jf with the spectra of SN~2016gkg at +320 d and +419 d, around the [O~I] and [Ca~II] doublets. All spectra are plotted in linear flux scale. The resolutions of the spectra at [O~I] are indicated.}
}
\label{narrowl}
\end{figure*}

Furthermore, SN~2016gkg displays a very prominent narrow component clearly in each line of the [O~I] doublet at $\lambda\lambda$6300 and 6364, which is not seen in other SNe thus far published.
The type-Ib SN~2009jf \citep{valenti11,sahu11} is possibly the closest analog to SN~2016gkg in terms of narrow [O~I] emission as its nebular spectra show a strong narrow component superposed on the broad [O~I] line, although without clear doublet identifications. The single narrow line in SN~2009jf is accompanied by weaker peaks in both blue and red directions, and altogether the [O~I] structure is interpreted as having originated from a number of clumps superposed on the bulk high-velocity ejecta along the line of sight.
{Figure~\ref{narrowl} shows the comparison between the spectra of SN~2016gkg and SN~2009jf, at photospheric and nebular phases. While the narrow [O~I] $\lambda$6300 is clearly evident in SN~2009jf, the $\lambda$6364 component is considerably weaker and ambiguous with the other smaller peaks. The narrow components in [Ca~II] and [Fe~II] are not visible. As the +361~d spectrum of SN~2009jf and the +320~d spectrum of SN~2016gkg have a similar resolution, the presence of those narrow components in SN~2009jf can therefore be ruled out.}

{The narrow [O~I] line in SN~2016gkg possibly appeared as early as three months after the light curve peak. Figure~\ref{narrowl} shows the $\sim$~+90 and +400~d spectra of SN~2016gkg, as compared to SN~2009jf that also shows a narrow [O~I] line core. Similarly, SN~2009jf also shows the emergence of this narrow line at around three months post-maximum. At this epoch, there is also an indication of the narrow core in O~I $\lambda$7774 in SN~2016gkg, however this is not clearly seen in SN~2009jf. The narrow cores in other lines (e.g. [Ca~II] and [Fe~II]) do not appear to be present in the early phase.
Note that in SESNe the ejecta become optically thin in the continuum relatively early -- generally within a few weeks after the light curve maximum. Therefore, conditions are nebular much sooner (i.e. the full ejecta is visible); forbidden lines simply require a longer time to become stronger.}

\citet{matheson00} noted that SN~1993J showed small-scale structures at the top of the O and Mg lines, which were interpreted as multiple clumps with distinct velocities. The Ca lines of SN~1993J however do not show similar profiles, and therefore it was interpreted that the Ca distribution in the ejecta of SN~1993J is more uniform. 
{For comparison, SN~2009jf has been claimed to have a more mixed ejecta due to the similar profile seen in the Mg~I] and [Ca~II] lines modified with an additional doublet component}. 
This is likely to be the case for SN~2016gkg as well, since the similar broad line + narrow core profile is also clearly seen in [Ca~II] (Figure~\ref{specvel}).
Ca is the strongest coolant in SE SNe, and thus [Ca~II] emission may arise from all regions where Ca is present. In contrast, [O~I] may only emit from the layers where O is the most abundant species.

\begin{figure}
   \centering
   \includegraphics[width=\hsize]{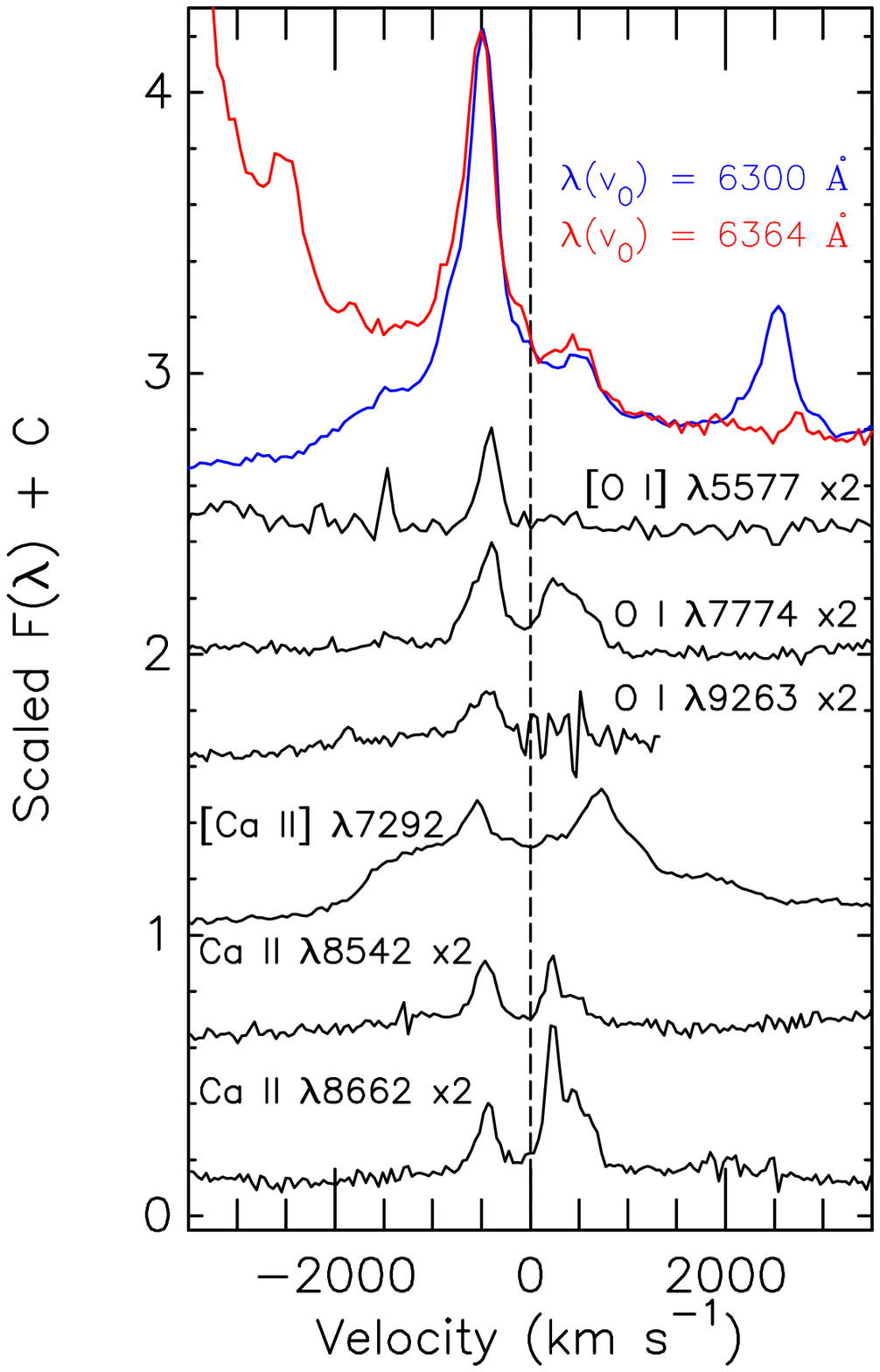}
\caption{Close up of narrow emission lines, in velocity space, in the MUSE spectrum at +419 d. Dashed vertical line indicates zero velocity for each of the indicated lines. The lines are scaled in flux if indicated ($\times$2), and in the case of [O~I] $\lambda$6364 (plotted in red) it is scaled to match the $\lambda$6300 component (plotted in blue). The profiles of these mirrored narrow components at $\pm$500 km~s$^{-1}$ match with each other well.
}
\label{specOCa}
\end{figure}

{Careful examination of the +419~d spectrum of SN~2016gkg shows that a blueshift is present in most of the emission lines (Figures~\ref{specvel} and \ref{specOCa}).}
In addition to the two most dominant doublets [O~I] $\lambda\lambda$6300,6364 and [Ca~II] $\lambda\lambda$7292,7324, it is also present in the other species such as Mg~I] $\lambda$4571 (Figure~\ref{specvel}), [O~I] $\lambda5577$, O~I $\lambda$7774, O~I $\lambda$9263, and also in the Ca~II lines of $\lambda$8542 and $\lambda$8662 (Figure~\ref{specOCa}). 
These lines do not appear to have the broad components. As the [O~I] $\lambda5577$, O~I $\lambda$7774, O~I $\lambda$9263, and Ca~II $\lambda$8542 and $\lambda$8662 lines are associated with high-density regions, this suggests that the narrow line components originated from the inner ejecta.
Accompanying the blue component, the red component is visible also in O~I $\lambda$7774, Ca~II $\lambda$8542, and Ca~II $\lambda$8662 {(Figure \ref{specOCa})}, and as previously discussed in the [O~I] $\lambda\lambda$6300,6364 doublet. It is possible that such structure is also present in [Ca~II] $\lambda\lambda$ 7292,7324, however since the narrow red components are considerably fainter relative to the blue ones they are not readily visible.

The nebular spectra of SESNe usually show broad emission lines of [O~I] and [Ca~II], with FWHM of around a few 1000 km~s$^{-1}$ \citep{tauben09}. 
This occurs because the energy deposition due to $\gamma$-rays {prevails} non-locally and thus {excites} the emission lines over a large volume (i.e. velocities, given homologous expansion). SN~2016gkg shows this typical line-width characteristics coexisting with narrow emission lines.
To produce narrow emission lines, a certain amount of material at low velocities needs to be powered by some source.

We note here that the identification of the narrow components in SN~2016gkg is not because of the relatively high spectral resolution of MUSE ($R\sim2200$) compared to the typical resolutions ($R\sim$ 500-1000) at which nebular spectra of SNe are usually acquired. In fact, they are clearly evident in the low-resolution GMOS and FOCAS spectra ($R\sim700$) as well {(Figures \ref{specall} and \ref{narrowl})}. While the MUSE resolution is not needed to identify the narrow lines, it is clearly required in order to resolve them.
Higher resolution does help in detecting narrow emission lines; regardless, such lines as in SN~2016gkg would have been detected if they were present in the other SNe observed thus far.
Therefore the absence of such narrow component in other SNe cannot be attributed to the low spectral resolution, and the fact that it is detectable and prominent in SN~2016gkg suggests that there must be a unique circumstance in this SN that leads to the emergence of the line.

\begin{figure}
   \centering
   \includegraphics[width=\hsize]{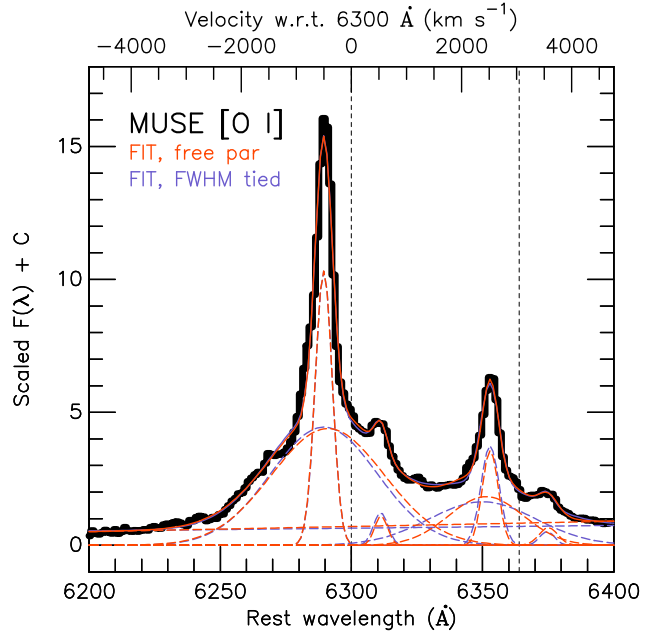}
\caption{Fitting of the [O~I] $\lambda\lambda$6300,6364 doublet. The observed spectrum is plotted in black continuous line, while the fitted functions in long-dashed lines (red for the free-parameter fit and blue for the FWHM-tied fit).
Short-dashed vertical lines indicate zero velocities for 6300 \AA~and 6364 \AA~rest wavelengths.}
\label{specgau}
\end{figure}

To measure the individual components comprising the [O~I] doublet complex, spectral line decomposition was performed. This was done using MPFIT \citep{markwardt09}, via the PAN: Peak Analysis application\footnote{\url{https://www.ncnr.nist.gov/staff/dimeo/panweb/pan.html}}. The MUSE spectrum at +419~d, corrected for the host redshift, was used due to the superior spectral resolution compared to the other spectra. The fit assumed a linear background and multiple Gaussian profiles in order to reproduce the observed line profile. Examining the region around the [O~I] line, it appears that there are three components in the [O~I] $\lambda$6300 line: a broad component, a narrow component blueward ($\sim$10~\AA~offset) of the rest wavelength of 6300~\AA, and a narrow component redward of 6300~\AA~(also with $\sim$10~\AA~offset). A similar profile is seen in the weaker [O~I] $\lambda6364$ line. Using the apparent line centers, peak intensities, and FWHM as initial guesses, we fitted six Gaussians and a linear background function to the [O~I] doublet line profile, without fixing any parameters. As such, the line centers, widths, and intensities, are left as free parameters. The result of the fit is shown in Figure~\ref{specgau} and the parameters are listed in Table~\ref{tabfit}.

\begin{figure*}
   \centering
   \includegraphics[width=\hsize]{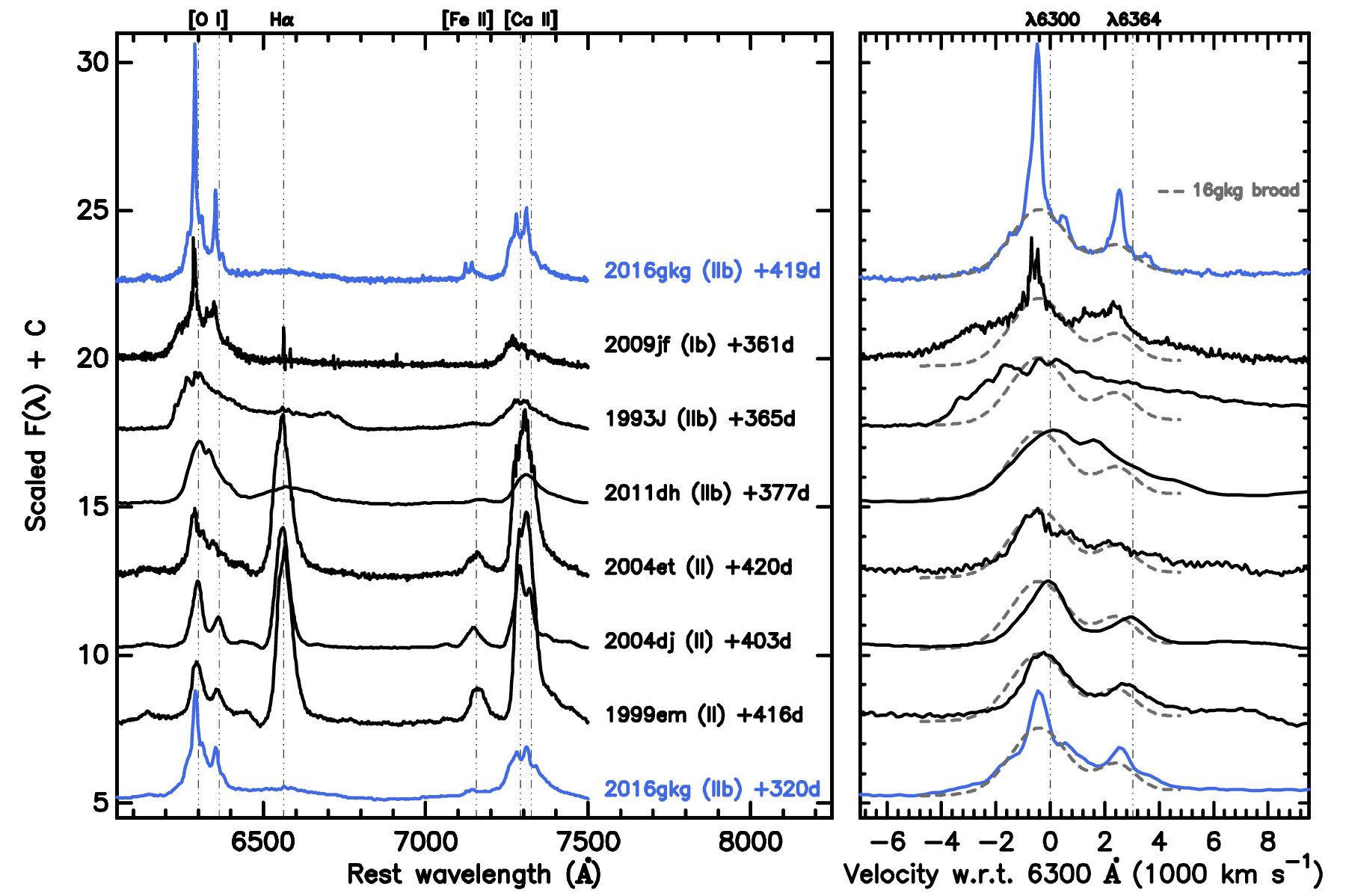}
\caption{{(Left)} Spectra SN~2016gkg compared to other SNe representative of the SN Ib/IIb and II classes, at similar phases: SN~2009jf \citep[Ib;][]{valenti11}, SN~1993J \citep[IIb;][]{matheson00}, SN~2011dh \citep[IIb;][]{ergon15}, SN~2004et \citep[II;][]{sahu06}, SN~2004dj \citep[II;][]{silverman17}, and SN~1999em \citep[II;][]{faran14}. The most prominent lines are indicated. Spectra are scaled to match the broad [O~I] components and shifted for clarity.
{(Right)} [O~I] profile comparison (zoom-in of left panel), with the broad component of SN~2016gkg shown by dashed lines. SN~2016gkg appears more similar to SNe II in terms of the broad component profile and width.
}
\label{speccomp}
\end{figure*}

Comparing with other stripped-envelope SNe at similar epochs, the broad component in SN~2016gkg appears to be narrower, while looking more similar to SNe~II in both profile and width (Figure~\ref{speccomp}). \citet{tauben09} reported typical FWHM of the [O~I] doublet from single-component Gaussian fit for SNe~Ib/IIb to be $\sim5200$ km~s$^{-1}$, while \citet{silverman17} noted that SNe~II typically show [O~I] half-width at half-maximum (HWHM) of 1000-1200 km~s$^{-1}$ (FWHM $\sim$ 2000-2400 km~s$^{-1}$). The broad components of SN~2016gkg are observed to have FWHM of $\sim2000$ km~s$^{-1}$, while the narrow components FWHM are $\sim300$ km~s$^{-1}$.
These are well greater than the instrument resolution of 140 km~s$^{-1}$, and thus correcting for it would not significantly reduce the derived FWHM velocities.
The doublet nature of the [O~I] emission line is more easily discerned in SNe II, whereas in SESNe the doublet is usually blended due to the higher expansion velocity. SN~2016gkg, while belonging to the SESN group, clearly displays the doublet nature of [O~I].
Difference in the broad component probably means that SN~2016gkg differs from other SESNe {in terms of} different inner ejecta structure, which may be characterized by higher density overall, or more $^{56}$Ni at low velocity.

%09jf: 362d, assuming tmax = 55119 (Valenti+11, Sahu+11) modjaz+14 typo!!!

\begin{deluxetable*}{lccccc}
\tablecaption{Resulting parameters of the [O I] doublet fit, +419 d spectrum. The spectral resolution is $\sim3$ \AA~$\approx 140$ km~s$^{-1}$; nominal fitting uncertainties are typically $0.1$ \AA~ $\approx 5$ km~s$^{-1}$. 
}
\tablewidth{0pt}
\tablehead{
\colhead{Component} & \colhead{$\lambda_\mathrm{center}$} & \colhead{$v_\mathrm{center}$} & \colhead{{FWHM}} & \colhead{$v_\mathrm{FWHM}$} & \colhead{Normalized flux} \\
\colhead{ } & \colhead{[\AA]} & \colhead{[km~s$^{-1}$]} & \colhead{[\AA]} & \colhead{[km~s$^{-1}$]}  & \colhead{ }
}
\startdata
Broad-6300 & 6291 & $-428$ & 50 & 2381 & 1.00* \\
Broad-6364 & 6351 & $-613$ & 34 & 1603 & 0.28 \\
Narrow-6300, blue & 6289 & $-524$ & 8 & 381 & 0.36 \\
Narrow-6300, red & 6311 & +524 & 6 & 286 & 0.03 \\
Narrow-6364, blue & 6353 & $-471$ & 7 & 330 & 0.12 \\
Narrow-6364, red ~~~~~ ~~~~~ & 6375 & +519 & 8 & 377 & 0.03 \\
\enddata
\tablecomments{*Equals to $2.36 \times 10^{-15}$ erg~s$^{-1}$~cm$^{-2}$~\AA$^{-1}$. }
\label{tabfit}
\end{deluxetable*}

\subsection{Properties of the oxygen-rich region}

Here we analyze the core oxygen properties based on the [O~I] doublet complex, using the spectrum at +419~d. The measured line ratio between the 6300~\AA~and 6364~\AA~lines yields 3.6$\pm$0.1~:~1 for the broad components, while the ratio is 2.8$\pm$0.1~:~1 for the narrow components (blue+red combined), suggesting a relatively denser narrow-line emitting region as the ratio should be close to 3~:~1 in the optically thin regime \citep{li92,jerkstrand17}, and 1~:~1 in the optically thick regime.
The [O~I] doublet profile in late-time SESNe has been discussed in the literature \citep{maeda08,tauben09}, in particular whether the double-peak features result from the $\lambda\lambda$6300,6364 doublet components or distinct velocity components in the ejecta \citep{milisav10}. Here in SN~2016gkg, the [O~I] line profile with the 64 \AA~peak separation and close to 3~:~1 flux ratio unambiguously point to the doublet origin.

The line ratio of $\lambda$6300 and $\lambda6364$ (combining the broad and narrow components) can be compared with theoretical calculations to derive the density of the [O~I] emitting region.
Using Figure~6 of \citet{leibundgut91}, these line ratios yield an {estimated} [O~I] density of $<3\times10^8$ cm$^{-3}$. The [O~I] $\lambda5577$ line is detected in the spectrum, and thus can be used to estimate the temperature of the emitting region \citep[e.g.][]{elmhamdi11}. 
In SN~2016gkg, the observed [O~I]$\lambda$5577/($\lambda\lambda$6300+6364) line ratio is 0.07. Using equation (2) of \citet{jerkstrand14}, and assuming $\beta_{5577} / \beta_{6300,6364} = 1.5$, the estimated temperature is then 3800 K.
We adopt this temperature estimate and further derive the mass of the emitting oxygen from the total luminosity of the [O~I] line (accounting for both the narrow and broad components), using equation (3) of \citet{jerkstrand14} with $\beta_{6300,6364} = 0.5$.
{Note that the $\lambda$5577 line falls out of local thermodynamic equilibrium (LTE) conditions after around 250~d in SNe IIP models, and after 150~d in SN IIb models \citep{jerkstrand15}, although in SN~2016gkg this appears to be delayed due to the lower expansion velocity.
Even if $\lambda$5577 is in NLTE but $\lambda$6300,6364 still in LTE the method will give an upper limit to the oxygen mass, which is still useful.
} 

Assuming a distance of 26.4 Mpc and reddening $A_V = 0.053$~mag \citep{bersten18}, the total oxygen mass is therefore estimated to be in the order of 0.3 $M_\odot$, which accounts for the $\sim 0.2$ and 0.1 $M_\odot$ associated with the broad and narrow components, respectively, assuming that line flux corresponds to mass in the same way for both components.

The core oxygen mass is related to the initial mass of the progenitor star. Following the method presented in \citet{hk15}, the initial mass of the progenitor of the SN~2016gkg is estimated by comparing the core oxygen mass to the theoretical yields from \citet{nomoto97} and \citet{limongi03}. The resulting estimate for the progenitor initial mass with 0.3 $M_\odot$ of core oxygen mass is 13-15 $M_\odot$. 
This estimate is effectively a lower limit since some of the oxygen may not be emissive. 
{In such a case, for example assuming that half of the oxygen is not emissive, doubling the oxygen mass estimate of SN~2016gkg would bring the initial mass estimate to be around 16-17 $M_\odot$. }
In comparison, \citet{bersten18} suggested that the progenitor of SN~2016gkg might have been a 19.5 $M_\odot$ primary star in a close binary system, based on their {HST} progenitor detection and binary star evolution modeling. Whereas, \citet{kilpatrick17} suggested the mass to be around 15 $M_\odot$, and \citet{tartaglia17} estimated 15-20 $M_\odot$, from photometric analyses of the progenitor star detected in pre-explosion archival HST images.
%In 19-20 M$_\odot$ progenitor models half the oxygen would have to be too cool to make significant [O I] emission \citep{jerkstrand15}, and doubling the oxygen mass estimate of SN~2016gkg would bring the initial mass estimate to be around 20 $M_\odot$. 

We further estimate the oxygen filling factor in the ejecta following the method of  \citet{leibundgut91}. At 400 d, assuming an  isotropic expansion with constant velocity of 2000 km s$^{-1}$, the ejecta would fill a sphere with a volume of $1.4 \times 10^{48}$ cm$^3$. If this sphere was uniformly filled with oxygen with an upper limit of mean density of $3 \times 10^8$ atoms~cm$^{-3}$ as previously estimated, a total of around 5.5~$M_\odot$ of oxygen would be contained. As the estimated oxygen mass of 0.3 $M_\odot$ is far below this total oxygen mass ($\sim5 \%$; however this is a lower limit), the interpretation would be that the oxygen filling factor in the ejecta is probably low, or that a significant fraction of the oxygen is located at a density lower than $3 \times 10^8$ atoms~cm$^{-3}$.

\subsection{Two-component ejecta and explosion geometry}

In SN~2016gkg, the narrow [O~I] lines are found to be $\sim$ 300 km~s$^{-1}$ wide, indicating that the emitting region is expanding at least five to six times slower compared to the broad component ($\sim$ 2000 km~s$^{-1}$), and have a combined [O~I] flux ($\lambda\lambda6300 + 6364$) about 40\% of the broad component {(see Table \ref{tabfit})}. 
If the emitting region is assumed to be of constant density, the density is scaled by the mass and the inverse cube of velocity, where the mass of the material is proportional to the emission line flux.
{Considering} the line flux of the narrow [O~I] component is {about 40\%  of the broad component, and the expansion velocity 20\% of the broad component, the density} would be $0.4/(0.2)^3 = 50$ times higher compared to the broad component. In general (i.e. in central explosions such as core-collapse SNe), density drops with increasing velocity and thus the broad component of [O~I] should form at lower density than the narrow component. 
If {the absorbing material is} not dense, the power absorbed by the low-velocity material would raise the ionization significantly. Furthermore, a low-density inner region would not efficiently absorb power.

The above illustration is consistent with an interpretation of line profiles seen in other SNe, where a narrow core on top of a broader component indicates enhanced central density. {Further}, the central dense part of the ejecta may take the form of a torus-like structure, which when viewed from the side gives rise to the double peak profile symmetric around the rest wavelength \citep{tauben09}. In principle, a bipolar structure pointing towards and away from the observer may also give rise to the blue and red pair of narrow emission lines.
The line displacement from rest wavelength immediately suggests asymmetric ejecta, as in the spherical case no offset is expected.

Note that as mentioned earlier in Section~\ref{sec:obs}, the SN rest frame is assumed to coincide with the underlying H~II region. If the rest wavelength is instead the central wavelength of the broad [O~I]$\lambda$6300 emission line, then the interpretation would become more general: the broad and narrow components of [O~I] lines are mostly symmetric with a minor red clump, suggesting a dense core in spherical ejecta (although an inner disk-like structure viewed face-on may in principle also be accommodated in this scenario).

\begin{figure}
   \centering
   \includegraphics[width=\hsize]{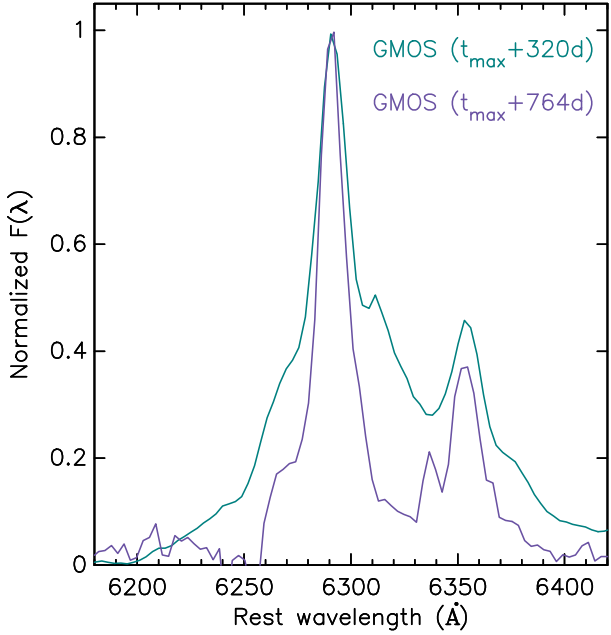}
\caption{Comparison of the two GMOS spectra with identical spectral resolution, at phases +320~d and +764~d at the [O~I] doublet region. The spectra are normalized in flux to match the peak intensity of the $\lambda$6300 line.}
\label{specgmos}
\end{figure}

Comparing the two GMOS spectra obtained with the identical instrument and resolution (+320~d and +764~d), the broad component is found to be diminishing in time with respect to the narrow component (Figure~\ref{specgmos}). 
One explanation might be that $\gamma$-rays increasingly leak from the ejecta, and thus the inner denser regions trap more efficiently relative to the faster-moving material above. The flux in the broad component then decreases faster than that of the narrow component.

Assuming that the broad component is emitted from a region defined by 0.2~$M_\odot$ and 2000~km~s$^{-1}$, then the $\gamma$-ray optical depth is described by the following: $\tau \approx 2\times10^4 \times (M/0.5M_\odot) \times (v/ \textrm{2000 km~s}^{-1})^{-2} \times (t/\textrm{day})^{-2}$. 
This component starts to become transparent to $\gamma$-rays (below $\tau \approx 1$) at around 90 days. After that, the luminosity would follow the decrease in the deposition rate, i.e., $\tau \propto t^{-2}$.
On the other hand, if the narrow component is defined by 0.1~$M_\odot$
and 300~km~s$^{-1}$, then $\tau$ is larger by a factor of 20 at around 90 days. For this component, the transition time to $\gamma$-ray transparency would be around 420 days. As a consequence, the luminosity ratio of the broad/narrow component would decrease over time, and after the narrow component transitions to be $\gamma$-ray thin, this ratio would saturate. 
This estimate, while rough, supports qualitatively that the narrow component is relatively denser and thus its transition to the optically thin phase is delayed. Such a phenomenon was predicted by the two-component SN explosion model of \citet{maeda03}.
Alternatively, as density increases inwards in all core-collapse SN explosions, the trapping of $\gamma$-rays is more efficient at low velocity than high velocity even in 1D spherical models. 

\begin{figure}
   \centering
   \includegraphics[width=\hsize]{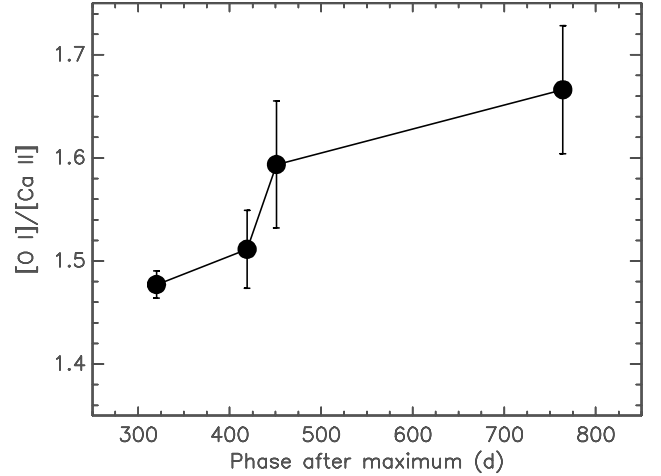}
\caption{Evolution of [O~I]/[Ca~II] flux ratio in SN~2016gkg, including both the broad and narrow components, integrated over the velocity range of $\pm$2500 km~s$^{-1}$.}
\label{ocaratio}
\end{figure}

The flux ratio of [O~I]/[Ca~II] has been frequently used as an indicator of core mass (see applications in e.g. \citealt{hk15}; initially in \citealt{fransson89}).
Modeling of SESN nebular spectra suggests that this ratio increases slowly with time \citep{jerkstrand17}.
In SN~2016gkg, the line ratio is found to be increasing (Figure~\ref{ocaratio}). 
The ratio increase might indicate that some part of the oxygen-rich region is not becoming transparent to $\gamma$-rays after around two years. At late time, a high density oxygen-rich core selectively absorbs $\gamma$-ray energy input. If the core is described by low-density and high-density regions (i.e. not a single component, \citealt{maeda03}), the latter would stay opaque to gamma rays and could emit for longer time. This behavior is seen in the diminishing broad component relative to the narrow component of SN~2016gkg (Figure~\ref{specgmos}).

The two-component ejecta model has been proposed in the past to describe the late-time light curves of hypernovae \citep{maeda03}, as a single-component model is insufficient to explain the observed late-time light curve and spectra of these objects \citep{sollerman00, mazzali00, mazzali01,izzo19}. 
This model describes the ejecta as consisting of two zones: the outer component that is dominating the early-phase light curve and the broad spectral features, and the dense inner component that is dominating in the later phase.
In the case of hypernovae, the presence of the dense core material has been predicted by jet-driven explosion models \citep[e.g.][]{maeda02}.
Such two-component shell model of the ejecta is however impossible to produce in 1D hydrodynamic simulations, and the only way to produce this is to break spherical symmetry \citep{dessart17}.

It has been established that {most, if not all,} stripped-envelope SN are aspherical explosions \citep{maeda08,tauben09}. The degree of asphericity generally range from extreme (explosions with relativistic jets, such as those of gamma-ray bursts and broad-lined type-Ic SNe) to mildly deviating from spherical geometry. In between, SN explosions may exhibit bipolar outflows \citep{maedanomoto03}. In a highly asymmetric explosion, the line of sight orientation may affect the observed nebular spectra. The broad-lined type-Ic SNe 1998bw and 2003jd are both inferred to be energetic explosions that launched relativistic jets. The jet was pointed towards Earth in the case of SN 1998bw, resulting in the gamma-ray burst (GRB) 980425, and also evident in the nebular spectra where the [O~I] line is sharp-peaked, and [Fe~II] is found at a higher velocity compared to [O~I] \citep{maeda02}. 

In a spherical explosion, oxygen is expected to be located in the outer layers relative to iron, and thus to show higher velocities. Whereas, in an aspherical event, explosive nucleosynthesis near the jet axis produces iron that moves with higher velocity compared to oxygen, which is distributed closer to the equatorial region, perpendicular to the jet axis.
Therefore, in the case of SN~2003jd, the line of sight goes through the jet equator as evidenced by the double-peaked nebular [O~I] line, and since the jet is pointing away from Earth, no GRB was observed \citep{mazzali05}. 
In SN~2016gkg, the double-peaked [O~I] line indicates that the line of sight is probing the oxygen-rich region that produces the strong [O~I] narrow emission.
However, there is no low-velocity analog to a jet such as those seen in hypernovae. The asymmetry in SN~2016gkg has to arise from something completely different.

Evidence of asymmetry or bipolarity in core-collapse SNe, in addition to nebular spectroscopy, include a mildly relativistic radio jet in the energetically-normal type-Ic SN 2007gr \citep{paragi10}.
SN~2007gr is also found to match the two-component explosion model, along with a number of type-Ib SNe \citep{cano14}.
In addition, observations of SN remnants also support this, such as the Si-rich bilateral jet of ejecta in the Vela SN remnant \citep{garcia17} and the distribution of $^{44}$Ti in the SN remnant Cassiopeia A \citep{grefenstette14}. 
SN light echo observations that probe a SN from multiple points of view, also add more evidence for explosion asymmetry \citep{rest11}.
In the case of hydrogen-rich (type-II) SNe, signs of bipolarity and asphericity have been observed (e.g. SN~2010jp, \citealt{smith12}; SN~2016X, \citealt{bose19}; SN~2017gmr, \citealt{andrews19,nagao19}), and it is likely that most of them are asymmetric explosions \citep{leonard06,leonard16}.
Even in low-energy SN II explosions such as in SN~2008bk, asymmetry is clearly detected \citep{leonard12}, and this further suggests that asymmetry does not require high energy explosions.
Since SESNe lack the hydrogen envelope, they should be even more asymmetric compared to the type-II SNe.
{Three-dimensional} simulations of neutrino-driven core collapse SN explosions also suggest that asymmetry is widespread \citep[e.g.][]{kifonidis00,kifonidis03,wongwa15}.

The velocity contrast between the narrow and broad components  in SN~2016gkg is very large, suggesting that two different regions are responsible for the emitted profiles. If the energy sources are the same (i.e. $^{56}$Ni), since the narrow component region encompass a much smaller volume (20\% of the broad component's velocity, therefore 0.8\% of the volume) then the factor 40\% in flux is an order of magnitude larger compared to the case where the oxygen density is homogeneous.
SN~2016gkg has been modeled as a standard-energy explosion with a kinetic energy of $1.2 \times 10^{51}$ erg and an ejecta mass of 3.4 $M_\odot$ \citep{bersten18}---in SESNe, where typical explosions produce $\sim$0.25--3.0 $\times 10^{51}$ erg of kinetic energy and 1--6 $M_\odot$ of ejecta \citep{lyman16,taddia18}, the ejecta material would be expelled with velocity in the order of thousands km~s$^{-1}$, and thus one would not expect material at low velocities. This material with low velocity is not predicted in 1D explosion models and immediately implies an asymmetric explosion.

SN~2016gkg thus presents a compelling evidence that significant asymmetry may be present even in a standard-energy explosion. The asymmetry is most striking at low velocity, possibly the lowest velocity material ever identified in a SESN, and not at large velocity. 
{Despite the presence of asymmetry, the SN explosion parameters, such as the explosion energy, ejecta mass, and $^{56}$Ni mass, inferred from modeling of the early phase (around light curve peak) with 1D explosion models would still hold as low-velocity material mostly affects the late-time light curves \citep{maeda03,maeda06} and spectra, as shown here.
In 1D explosion models, the energy needed to explain the early light curve and spectra would yield no material at low velocity \citep[see e.g.][]{dessart17}.
If multi-dimensional models are invoked, there could be mass at low velocity that makes the ejecta mass estimate from hydrodynamic simulations of the light curves to be slightly underestimated. This material would still have little impact to the derived explosion energy, due to its low velocity, and to the total ejecta mass, due to its small contribution ($\sim$0.1 $M_\odot$ in the case of SN~2016gkg).
In asymmetric neutrino-driven explosions, one typically observes that the explosion occurs in some directions, while continued accretion (and fallback) occurs in other directions. Low-velocity material would correspond to the material that just barely avoids fallback, and stays at low velocity in the final 3D ejecta \citep[e.g.][]{wongwa13,chan20}. Such material resides in the deeper layers of the exploding star and thus would not significantly affect the early light curves and spectra.
}

\subsection{Alternative interpretations of the narrow emission lines}

The narrow lines of SN~2016gkg cannot be explained by interaction with a circumstellar material (CSM). To produce a spectrum with strong, narrow [O~I] lines, the CSM needs to be oxygen-rich, and it has to be depleted in H and He as the narrow lines of these elements are not present. 
SN~2016gkg clearly showed signatures of H and He in the early phase \citep{tartaglia17,bersten18}, and thus an oxygen-rich CSM poor in H/He would require a fine tuning and is therefore highly unlikely.

Thus far, there is only one instance of a SESN interacting with H/He-free CSM in the H/He-poor type-Ic SN~2010mb \citep{benami14}, while on the other hand a type-Ic SN interacting with H/He-rich CSM displays strong H and He emission lines \citep[SN~2017dio,][]{hk18b}. 
The emergence of a blue continuum in SN~2010mb is also attributed to the interaction with the H/He-free CSM, and this feature is not apparent in SN~2016gkg.
In addition, the type-Ibn SNe characteristically display strong He emission lines resulting from interaction with He-rich CSM \citep[e.g.][]{pastorello07}. None of these examples show the characteristics seen in the spectra of SN~2016gkg.

\begin{figure}
   \centering
   \includegraphics[width=\hsize]{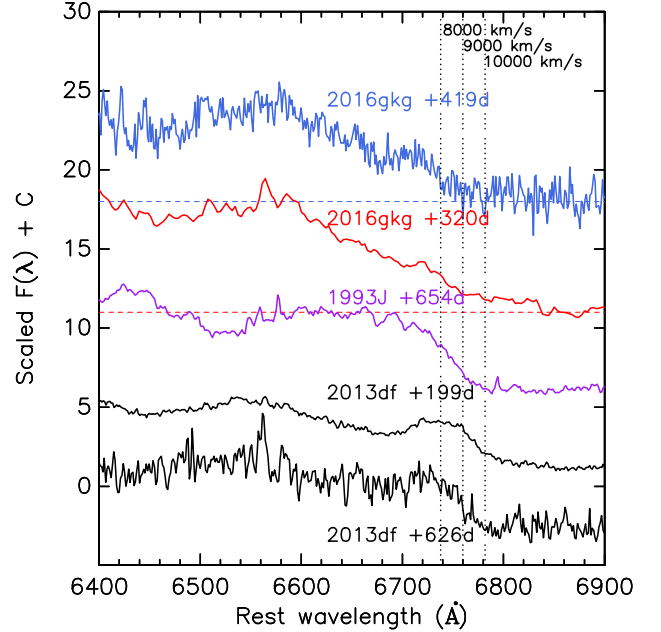}
\caption{Zoom-in on the H$\alpha$ region in the spectra of SN~2016gkg and other well-studied type-IIb SNe with CSM interaction.}
\label{specha}
\end{figure}

Weak interaction with H-rich CSM in SN~2016gkg is not ruled out, nevertheless, as the bump seen in the spectra around the H$\alpha$ region may be contributed by such interaction in addition to the emission from [N~II]. Such interaction with H-rich CSM has been seen in a number of SNe IIb, such as SNe 1993J and 2013df \citep{matheson00,maeda15}, {where} shock-induced H$\alpha$ emission may reach velocities around 10~000 km~s$^{-1}$ (see Figure~\ref{specha}). Still, this cannot explain the observed strong narrow lines in SN~2016gkg.

Another interesting possibility is the interaction with a stripped stellar companion. However, this scenario cannot explain the symmetric blue and red narrow line peaks.
The narrow lines of SN~2016gkg are also not likely to be the result from a pulsar wind nebula in the inner ejecta, as they are of low ionization. As exemplified by the type-Ib SN~2012au \citep[][]{milisav18}, the [O~III] $\lambda\lambda$4959,5007 and [O~II] $\lambda\lambda$7320,7330 appear strong, comparable to [O~I], as a result of a pulsar wind nebula inside the SN ejecta.
We note the absence of strong [O~II] and [O~III] emission lines, which could also arise from CSM interaction or from magnetar power \citep[][]{jerkstrand17sl,dessart19}, in the spectra of SN~2016gkg. [O~II] and [O~III] are more easily produced in low-density conditions, cf. [O~I] that is associated with high density.
There are weak features in the positions of [O~II] and [O~III] in the +764~d spectrum, therefore similarly a pulsar wind nebula may be present at the core, but its power contribution is limited even at the latest epoch and therefore it cannot be responsible for powering the set of conspicuous emission lines seen in the earlier spectra.

\subsection{SN environment}

NGC~613, the host galaxy of SN~2016gkg, is a large spiral galaxy with a measured metallicity 8\arcsec.6 north from the SN explosion site of 12+log(O/H) = 8.7 dex, close to the solar value \citep[][using a combination of indices]{bersten18}. 
The galaxy exhibits gas outflows near the center \citep{lopez20}, however this does not seem to affect the outer parts where SN~2016gkg lies.

We extracted the background stellar population of SN~2016gkg from our MUSE datacube, integrating light within an annulus extending between 0\arcsec.9-1\arcsec.7 ($\sim100$-200 pc) from the SN. 
We do not detect any anomalous flows around the SN in the H$\alpha$ velocity field extracted from the MUSE datacube. The redshift of the background narrow H$\alpha$ line is consistent with the surrounding H~II regions in the field and the host redshift.
Utilizing the [N~II]$\lambda$6584 and H$\alpha$ emission lines, an oxygen abundance of 12+log(O/H) = 8.6~dex in the N2 scale \citep{marino13} was calculated from the extracted stellar population spectrum, consistent with the near-solar metallicity estimate. 
The equivalent width of the H$\alpha$ emission line in this background stellar population was measured to be around 7~\AA. 
While other factors such as shocks and emissions from post-asymptotic giant branch stars may contribute to the observed emission line equivalent width \citep[see e.g.][]{cid11}, here we assume that the H$\alpha$ emission is contributed by the star-forming parent stellar population of the progenitor of SN~2016gkg and thus can be used as an age indicator.
Following \citet{hk18}, this H$\alpha$ equivalent width value is consistent with the stellar population age of around 12 Myr by comparing with Starburst99 simple stellar population models \citep{sb99}, which subsequently points to the lifetime of a single 16 $M_\odot$ star.
This estimate is consistent with the progenitor initial mass range of 15-20 $M_\odot$ derived with other methods discussed above.

\section{Summary}

We present late-time spectral observations of the type-IIb SN~2016gkg. The nebular spectra show the characteristic [O~I] $\lambda\lambda$6300,6364 and [Ca~II] $\lambda\lambda$7292,7324 emission doublets as the strongest lines, accompanied by other lines which are fainter. 

The most striking features of the spectra are the strong narrow lines superposed on the broad component base, reaching low velocities down to $\sim 300$ km~s$^{-1}$. The [O~I] doublet is the strongest line in the spectra, and both the $\lambda\lambda$6300 and 6364 lines feature a composite line profile consisting of at least three single components: a broad emission (FWHM $\sim$ 2000 km~s$^{-1}$), a blue and a red emission components mirrored against the zero velocity.
Such a low velocity seen in the narrow lines is not expected from spherical 1D SN explosion models, and points to the presence of slow-moving material in the ejecta. 

SN~2016gkg suggests that such low-velocity material might also be present in other SNe, although not observed thus far. 
A more detailed modeling of the power source at the origin of the narrow line component is needed.
While we made no attempt of modeling the explosion and spectra of SN~2016gkg, future SN modeling efforts may take advantage of these observations and incorporate more detailed ejecta structures to improve the constructed models of SNe and their progenitors. 

\begin{acknowledgements}
We thank the anonymous referee and Stefan Taubenberger for useful comments and discussions.
Archival data used in this article were obtained via the UC Berkeley Supernova Database\footnote{\url{http://heracles.astro.berkeley.edu/sndb/}} \citep{shivvers19}, the Open Supernova Catalog\footnote{\url{https://sne.space/}} \citep{guillochon17}, and the Weizmann Interactive Supernova Data Repository\footnote{\url{https://wiserep.weizmann.ac.il/}} \citep[WISeREP,][]{yaron12}.
H.K. was funded by the Academy of Finland projects 324504 and 328898.
K.M. acknowledges support by JSPS KAKENHI Grant (20H00174, 20H04737, 18H04585, 18H05223, 17H02864).
L.G. was funded by the European Union's Horizon 2020 research and innovation programme under the Marie Sk\l{}odowska-Curie grant agreement No. 839090. This work has been partially supported by the Spanish grant PGC2018-095317-B-C21 within the European Funds for Regional Development (FEDER).
C.P.G. acknowledges support from EU/FP7-ERC grant no. [615929].
F.O.E.\ acknowledges support from the FONDECYT grant nr.\ 11170953 and 1201223.
\end{acknowledgements}

{\facilities{Gemini-S (GMOS-S), Subaru (FOCAS), VLT (MUSE)}}

{\software{Astropy \citep{astropy18}, ESO Reflex \citep{freudling13}, GDL \citep{coulais10}, IRAF \citep{tody86,tody93}, L.A.Cosmic \citep{vandokkum01}, mpfit \citep{markwardt09}, QFitsView \citep{ott12}, ZAP \citep{soto16} 
}}


\end{document}